\documentclass[runningheads]{llncs}

\usepackage[T1]{fontenc}
\usepackage[utf8]{inputenc}
\usepackage{microtype}
\usepackage{graphicx}
\usepackage{xcolor}
\usepackage{tcolorbox}
\usepackage{tabularx}
\usepackage{booktabs}
\usepackage{makecell}
\usepackage{array}
\usepackage{float}
\usepackage[hidelinks]{hyperref}
\title{WebArxiv: A Reproducible Benchmark for Evaluating Multimodal Web Agents on arXiv Tasks}
\titlerunning{WebArxiv: A Benchmark for Multimodal Web Agents}
\author{Zihao Sun \and Zijing Shi \and Ling Chen}
\authorrunning{Z. Sun et al.}
\institute{Australian Artificial Intelligence Institute,\\
University of Technology Sydney, Sydney, Australia}
\newcounter{webarxivAppendixStart}
\newcounter{webarxivAppendixPages}
\newcounter{webarxivTotalPages}
\newcounter{webarxivMainPages}

\newcommand{\TaskFiftyOneScreenshot}[1]{%
  \begin{minipage}[c][0.17\textheight][c]{\linewidth}%
    \centering
    \includegraphics[width=\linewidth,height=0.17\textheight,keepaspectratio]{#1}%
  \end{minipage}%
}

\begin{document}
\maketitle

\begin{abstract}
Foundation models now enable autonomous agents to interact with real-world websites, but existing benchmarks emphasize general-purpose browsing, underrepresent research-oriented environments and scholarly discovery workflows, and often depend on live sites whose changing content and structure undermine reproducibility. arXiv provides a realistic, reproducible, hierarchically structured, information-centric testbed without privacy-sensitive interactions. We introduce WebArxiv, a static-snapshot benchmark comprising 510 time-invariant tasks, each with a unique deterministic ground truth. Its diverse, realistic scholarly tasks go beyond simple information lookup and rule following to emphasize multi-constraint paper retrieval, fine-grained content extraction, and cross-paper comparison. Evaluations of a range of foundation-model-based web agents show that WebArxiv remains challenging. Behavioral analysis reveals that agents over-rely on fixed interaction histories, causing incomplete or repetitive reasoning. We therefore equip agents with a lightweight dynamic-memory mechanism for adaptive retrieval and reasoning over relevant context. The benchmark and code are available at \url{https://anonymous.4open.science/r/74E4423BVNW/README.md}.

\end{abstract}

\section{Introduction}
Rapid advances in large language models (LLMs) and large multimodal models (LMMs) enable autonomous agents to perform increasingly complex tasks on real-world websites \cite{garg2025real,yang2025agentic}. Research-oriented agents must additionally interpret structured scholarly metadata, navigate hierarchical taxonomies, and support scholarly workflows. Yet existing evaluations emphasize broad assistant tasks \cite{mialon2023gaia} or general web browsing \cite{koh2024visualwebarena}; scientific benchmarks \cite{chen2024scienceagentbench,wang2025liveresearchbench} instead assess downstream scientific problem solving. Workflows on real research platforms therefore remain underexplored.

arXiv is an attractive testbed because it is stable and transparent, provides well-structured metadata, and has minimal personalization. WebVoyager \cite{he2024webvoyager} includes arXiv among 15 live websites, but its arXiv tasks mainly test basic navigation rather than scholarly workflows, and answers may change as live pages evolve. Its automatic GPT-4V evaluator uses model-based judgment rather than deterministic exact matching. These differences motivate a stable evaluation setting for scientific workflows.

We introduce WebArxiv, covering diverse, realistic scholarly tasks beyond basic information queries and rule compliance, including multi-condition paper retrieval, deep content extraction, and cross-paper comparison. Static arXiv snapshots make every task time-invariant, while reference action trajectories and deterministic ground truths enable consistent, verifiable comparisons among agents.

We evaluate ten representative configurations, spanning general-purpose LMM agents such as GPT-4o \cite{openai2024gpt4o} and specialized systems such as OpenWebAgent \cite{iong2024openwebagent}. The strongest model, GPT-o1, reaches 47.6\% overall success, followed by Gemini-2.5 at 44.4\%, while most models remain below 38.0\% and struggle especially with multi-condition advanced search and paper comparison. Qualitative trajectory inspection finds repeated actions after agents rely on short recent histories. Dynamic memory instead retrieves task-relevant prior interactions, helping avoid error loops and reuse information during retries; the experiments report improvements for nearly all evaluated models.

Our contributions are summarized as follows:
\begin{itemize}
\item We introduce WebArxiv, a benchmark that enables reproducible evaluation of multimodal web agents within the hierarchically structured arXiv environment.
\item We conduct a comprehensive evaluation of ten advanced web agents on WebArxiv, demonstrating clear baseline performance across different task types and query complexity levels.
\item Through detailed analyses, we find that existing web agents tend to over-rely on fixed interaction histories.
To address this issue, we introduce a lightweight dynamic memory mechanism that allows web agents to adaptively leverage past interactions.
\end{itemize}

\section{Related Work}

WebArxiv builds on web-agent evaluation but shifts attention to research tasks in real web environments.

\subsection{Benchmarks for Web Agents.}

Early LLM/LMM benchmarks \cite{liu2023agentbench} used simulations: WebShop models a virtual store with over one million Amazon listings \cite{yao2022webshop}, while WebArena spans simulated e-commerce, social-media, and productivity sites \cite{zhou2024webarena}. Later live-web benchmarks include Mind2Web, which provides 2,350 tasks across 137 websites \cite{deng2023mind2web}; BrowseComp, which targets complex comparison and open-domain browsing \cite{wei2025browsecomp}; and multimodal WebVoyager, which integrates text and vision across 15 real-world sites \cite{he2024webvoyager}. Although these benchmarks capture real-web unpredictability \cite{workarena2024}, evolving pages cause answer drift, reduce reproducibility, and require ground-truth updates \cite{chezelles2025browsergym}. WebVoyager and Mind2Web 2 therefore use automated model-as-a-judge outcome evaluation to reduce human annotation \cite{he2024webvoyager,gou2025mind2web}. They nevertheless cover research workflows sparsely \cite{wang2022selfinstruct}, motivating benchmarks that retain real-world complexity while providing stable ground truths \cite{levy2025stwebagentbench}.

\subsection{Web Agents.}
Modern LLM/LMM agents execute end-to-end user instructions through direct interaction with real websites \cite{ma2024agentboard,shi-etal-2026}. ReAct interleaves reasoning and actions \cite{yao2022react}; Reflexion and Auto-Eval and Refine analyze failures to iteratively improve decisions \cite{shinn2023reflexion,pan2024autonomous}; and WebPilot uses Monte Carlo Tree Search to evaluate, backtrack, and optimize trajectories \cite{zhang2025webpilot,koh2024visualwebarena}. Fine-tuning language or multimodal models provides another route \cite{pan2024webcanvas}: WebGPT tunes GPT for browsing and question answering \cite{nakano2023webgpt}, whereas HTML-pretrained WebAgent learns structural understanding and action grounding from large HTML corpora \cite{gur2024webagent,hong2024cogagent}. LMM agents instead use screenshots for holistic layout and content perception \cite{gou2025uground}. Most multimodal agents rely on closed-source LMMs such as GPT-4V and Gemini, whose visual grounding and reasoning support ReAct- and Reflexion-style planning \cite{zheng2024seeact,anil2023gemini,yao2022react,shinn2023reflexion}. Pix2Act and WebGUM map screenshots to grounded actions without explicit DOM parsing \cite{shaw2023pix2act,furuta2024webgum}; SeeAct combines LMM planning with HTML-and-visual grounding \cite{zheng2024seeact}.

\section{WebArxiv Benchmark}

\subsection{Overview}

\begin{figure}[!t]
\centering
\includegraphics[width=0.92\textwidth]{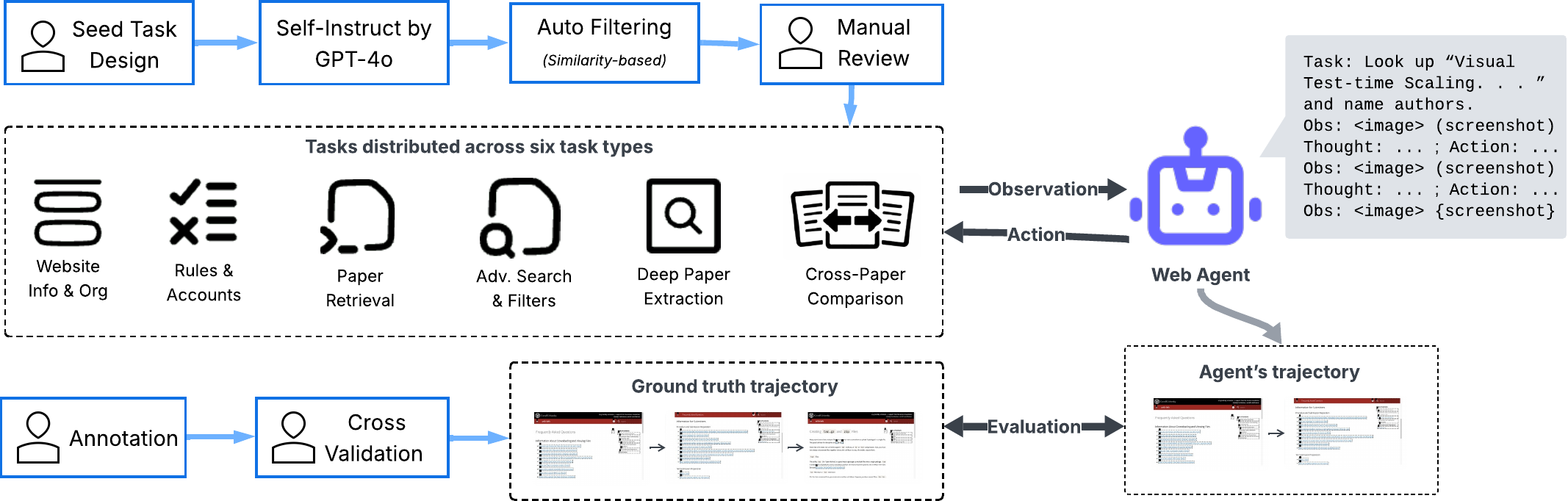}
\caption{WebArxiv construction pipeline. Seed tasks are expanded, similarity-filtered, manually reviewed, and expert cross-validated.}
\label{fig:mindmap}
\end{figure}

WebArxiv is a static-snapshot benchmark of 510 complex scholarly tasks designed for reproducible evaluation, including multi-condition paper retrieval, deep content extraction, and cross-paper comparison. Each task includes a natural-language instruction, reference action trajectory, and machine-verifiable ground truth validated by multiple annotators. Figure~\ref{fig:mindmap} summarizes its construction, and Table~\ref{tab:data-distribution} reports the distribution and detailed task statistics.

\begin{table}[!t]
\caption{Data distribution across six task categories. Counts summarize task generation and filtering; ground-truth steps come from the human-annotated reference trajectories.}
\label{tab:data-distribution}
\centering
\setlength{\tabcolsep}{6pt}
\resizebox{\linewidth}{!}{
\begin{tabular}{lcccccc|c}
\toprule
\textbf{Statistic / Category} &
\makecell{\textbf{Website}\\ \textbf{Info \& Org}} &
\makecell{\textbf{Rules \&}\\ \textbf{Accounts}} &
\makecell{\textbf{Paper}\\ \textbf{Retrieval}} &
\makecell{\textbf{Adv.\ Search}\\ \textbf{\& Filters}} &
\makecell{\textbf{Deep Paper}\\ \textbf{Extraction}} &
\makecell{\textbf{Cross-Paper}\\ \textbf{Comparison}} &
\textbf{Total} \\
\midrule
Generated tasks            & 100 & 100 & 200 & 200 & 200 & 200 & 1000\\
Similar tasks              & 34  & 37  & 31  & 51  & 28  & 44  & 225\\
Unable to annotate         & 5   & 5   & 7   & 11  & 32  & 38  & 98\\
Expired tasks              & 6   & 3   & 62  & 38  & 40  & 18   & 167\\
\textbf{Refined tasks}     & 55  & 55  & 100 & 100 & 100  & 100  & \textbf{510}\\
GT Annotate steps          & 270 & 274 & 639 & 752 & 884 & 1613 & 4432\\
Avg.\ GT steps/task        & 4.9 & 5.0 & 6.4 & 7.5 & 8.8 & 16.1 & 8.69\\

\bottomrule
\end{tabular}
}

\end{table}

\subsection{Problem Formulation}
Following prior work \cite{zhou2024webarena}, we formulate web interaction as a partially observable Markov decision process (POMDP) \cite{kaelbling1998planning}:
$\mathcal{E} = (\mathcal{S}, \mathcal{A}, \mathcal{T}, \mathcal{O}, \Omega),$
where $\mathcal{S}$ is the web state space, $\mathcal{A}$ contains user-interaction primitives, $\mathcal{T}$ is the transition function, $\mathcal{O}$ is the observation space, and $\Omega$ maps latent states to multimodal observations. Each action uses a template from Table~\ref{tab:agent_actions} and, where applicable, an accessibility-tree element ID.

At step $t$, the environment exposes $o_t \in \mathcal{O}$ as a rendered-page screenshot and accessibility tree, which provide complementary visual and semantic information; each evaluated system consumes the subset supported by its native interface. Given instruction $i$ and history $h_t$, its policy selects $a_t \in \mathcal{A}$ as $\pi(a_t \mid i,h_t)$. Execution transitions the environment to $s_{t+1}$ and returns $o_{t+1}$; the episode ends when the task is completed or the bounded step budget is exhausted.

\begin{table}[!t]
\caption{Available action space of the WebArxiv agent. The entries specify permitted interface actions rather than annotation-derived statistics.}
\label{tab:agent_actions}

\centering
\small
\setlength{\tabcolsep}{6pt}
\renewcommand{\arraystretch}{1.0}
\begin{tabularx}{\columnwidth}{@{}p{2.2cm} X@{}}
\toprule
\textbf{Action} & \textbf{Syntax} \\
\midrule
Click 
& \texttt{Click [Numerical\_Label]} \\
Type
& \texttt{Type [Numerical\_Label]; [Content]} \\
Scroll 
& \texttt{Scroll [WINDOW]; [up/down]} \\
Wait
& \texttt{Wait} \\
GoBack
& \texttt{GoBack} \\
Answer
& \texttt{ANSWER; [Your Answer]} \\
\bottomrule
\end{tabularx}

\end{table}

\subsection{Task Types}
WebArxiv contains 510 tasks in six categories, organized by information needs and interaction requirements. The category schemas used for seed-task creation and expansion are as follows:

\noindent\textbf{Website Information and Organizational Details (Website Info \& Org)} tests navigation and retrieval of editorial structures, platform descriptions, and organizational guidelines.\par
\noindent\textbf{Rules, Licensing, and User Account Management} tests procedural and policy retrieval for submission requirements, licenses, withdrawals, and account operations.\par
\noindent\textbf{Research Paper Discovery and Retrieval} tests search by author, paper ID, or subject and extraction of stable titles, authors, and primary subjects.\par
\noindent\textbf{Advanced Search and Filtering} tests multi-constraint structured queries combining keywords, publication years, and subject domains.\par
\noindent\textbf{Deep Paper Content Extraction} tests fine-grained textual and structural extraction from paper HTML.\par
\noindent\textbf{Cross-Paper Comparison} tests synthesis across papers' sections, experiments, and findings, evaluating cross-document reasoning.\par

Advanced Search and Filtering, Deep Paper Content Extraction, and Cross-Paper Comparison require the most interaction steps; Appendix~\ref{app:task} gives a worked task example.

\subsection{Task Construction Pipeline}

Experts wrote 20 seed tasks per category. Guided by category schemas and handcrafted templates, GPT-4o expanded them to 100 candidates for each simple category and 200 for each complex category, simulating realistic queries and interaction intents. We then filtered candidates in two stages: first, we removed semantic redundancy using pairwise sentence similarities from \texttt{all-\allowbreak{}mpnet-\allowbreak{}base-\allowbreak{}v2}, retaining scores below 0.8; second, the research team manually checked clarity and excluded tasks that were unreliable to annotate, had non-unique ground truths, or were not time-invariant. The resulting tasks form WebArxiv; Table~\ref{tab:data-distribution} summarizes their distribution.

\subsection{Benchmark Construction}
\label{sec:benchmark-construction}

To convert dynamic pages into stable, replayable environments with consistent multimodal observations, we use four stages.

\noindent\textbf{Snapshot Acquisition.} Selenium WebDriver captures HTML, accessibility trees, and full-page screenshots under Chrome v118.0 on Ubuntu 22.04 LTS at 1920×1080, providing consistent structural, semantic, and visual representations.\par
\noindent\textbf{Recursive Crawling and Caching.} From root arXiv category pages, we recursively crawl each target URL and cache its HTML, linked stylesheets, and embedded scripts for offline reconstruction and consistent replay.\par
\noindent\textbf{Deterministic Re-rendering.} A sandboxed browser with the same pinned rendering engine reduces run-to-run rendering variation and removes network and live-content changes. Because arXiv's core pages are server-generated and do not depend on client-side JavaScript, reconstruction remains faithful.\par
\noindent\textbf{Reproducibility Packaging.} Each task bundles its HTML snapshot, accessibility tree, full-page screenshot, hashed metadata, and JSON interaction trajectory, enabling deterministic replay of environment states and agent interactions across systems.\par

\subsection{Annotation Process}

Five PhD-level annotators experienced with arXiv manually annotated all 510 tasks, guided by one reference example per category. During initial annotation, each annotator independently completed assigned tasks in WebArxiv, recording every action and key screenshots through the final answer. A custom Playwright~v1.45 interface logged each click, scroll, and input to produce fully traceable reference trajectories.

During verification, annotators cross-validated every task for correctness, consistency, and agreement; the agreed final answer and its reference trajectory became the evaluation ground truth. We compare each model's final output with the reference answer and inspect screenshot-based action trajectories for logical consistency. Outcomes are \textbf{Correct} when the output satisfies the exact-match rule against the gold answer and \textbf{Incorrect} otherwise.

\subsection{Dynamic Memory.}

\begin{figure}[!t]
\centering
\begin{tcolorbox}[boxsep=0.4mm,left=1mm,right=1mm,top=0.4mm,bottom=0.4mm,fontupper=\small]
For each interaction step $t$:\par
1. Retrieve the full interaction history $\mathcal{H}_t=\{(o_1,a_1),\ldots,(o_t,a_t)\}$.\par
2. Ask the model to select relevant indices $J^*\subseteq\{1,\ldots,t\}$ with $|J^*|\le m$.\par
3. Construct dynamic memory $\mathcal{M}_t=\{o_j\mid j\in J^*\}$.\par
4. Combine the task, current view $o_{t+1}$, and $\mathcal{M}_t$.\par
5. Generate and execute $a_{t+1}$, then append the new interaction to $\mathcal{H}_t$.
\end{tcolorbox}
\caption{Pseudocode for dynamic memory with adaptive selection from the full interaction history.}
\label{fig:memory-pseudocode}
\end{figure}

In baseline evaluations, agents often over-rely on a truncated recent view and overlook earlier task-relevant information, leading to reasoning loops or incomplete answers. We therefore equip agents with the lightweight dynamic memory shown in Fig.~\ref{fig:memory-pseudocode}. At step $t$, the agent retrieves from the accumulated history $\mathcal{H}_t$ and adaptively selects a set of useful past indices $J^*$, rather than fixed trace-back steps. The resulting memory $\mathcal{M}_t$ is combined with the task instruction and current view to generate
$a_{t+1}\sim\pi(i,o_{t+1},\mathcal{M}_t)$.
This mechanism is intended to reduce repetitive loops and enable retries that reuse information acquired before a failure or redirection.

\section{Experiments}
Table~\ref{tab:WebArxiv-selected-models} reports category-level performance. We then describe the baseline agents, evaluation protocol, main results, ablations, and analysis.

\begin{table}[!t]
\caption{Success rates of web agents across the six WebArxiv task categories. Values are averaged over three runs.}
\label{tab:WebArxiv-selected-models}
\centering
\setlength{\tabcolsep}{3.5pt}
\renewcommand{\arraystretch}{1.00}
\resizebox{\linewidth}{!}{
\begin{tabular}{lcccccc|c}
\toprule
Web Agents &
\begin{tabular}[c]{@{}c@{}}Website\\ Info \& Org\end{tabular} &
\begin{tabular}[c]{@{}c@{}}Rules \&\\ Accounts\end{tabular} &
\begin{tabular}[c]{@{}c@{}}Paper\\ Retrieval\end{tabular} &
\begin{tabular}[c]{@{}c@{}}Adv. Search\\ \& Filters\end{tabular} &
\begin{tabular}[c]{@{}c@{}}Deep Paper\\ Extraction\end{tabular} &
\begin{tabular}[c]{@{}c@{}}Cross-Paper\\ Comparison\end{tabular} &
Total (\%) \\
\midrule
GPT-4-Turbo      & 44.2\% & 33.9\% & 47.3\% & 25.3\% & 31.3\% & 14.7\% & 31.7\% \\
GPT-4o           & 35.8\% & 29.7\% & 36.3\% & 24.3\% & 38.7\% & 17.7\% & 30.0\% \\
GPT-o1           & \textbf{73.3\%} & 50.3\% & \textbf{66.7\%} & 42.3\% & \textbf{45.3\%} & 20.7\% & \textbf{47.6\%} \\
GPT-4o-mini      & 52.1\% & 49.7\% & 56.3\% & 28.3\% & 34.3\% & 14.0\% & 37.1\% \\
Gemini-1.5-pro   & 47.9\% & 41.2\% & 53.3\% & 33.3\% & 38.3\% & 18.7\% & 37.8\% \\
Gemini-2.0       & 35.2\% & 28.5\% & 35.3\% & 23.3\% & 26.7\% & 10.7\% & 25.7\% \\
Gemini-2.5       & 63.0\% & \textbf{58.8\%} & 54.3\% & \textbf{46.3\%} & 36.3\% & \textbf{22.3\%} & 44.4\% \\
\midrule
SeeAct           & 26.7\% & 22.4\% & 25.3\% & 20.3\% & 25.3\% & 8.7\%  & 20.9\% \\
LiteWebAgent     & 46.1\% & 44.8\% & 44.3\% & 33.7\% & 44.3\% & 21.3\% & 38.0\% \\
OpenWebAgent     & 33.9\% & 38.8\% & 44.7\% & 33.3\% & 19.3\% & 7.7\%  & 28.4\% \\
\bottomrule
\end{tabular}
}

\end{table}

\subsection{Baselines}

\begin{table}[!t]
\caption{Aggregate success, efficiency, and trajectory consistency.}
\label{tab:web-agent-comparison}
\centering
\setlength{\tabcolsep}{4pt}
\renewcommand{\arraystretch}{1.00}
\resizebox{0.88\linewidth}{!}{%
\begin{tabular}{lccc}
\toprule
\textbf{Web Agent} & \textbf{Success Rate (\%)} & \textbf{Avg. Step Count} & \textbf{Trajectory Consistency $\uparrow$} \\
\midrule
\textbf{GPT-o1} & \textbf{47.6} & \textbf{8.6} & \textbf{0.82} \\
GPT-4-Turbo & 31.7 & 9.4 & 0.71 \\
GPT-4o & 30.0 & 9.2 & 0.68 \\
GPT-4o-mini & 37.1 & 8.8 & 0.75 \\
Gemini-1.5-pro & 37.8 & 8.9 & 0.74 \\
Gemini-2.0 & 25.7 & 9.7 & 0.66 \\
\textbf{Gemini-2.5} & \textbf{44.4} & \textbf{8.9} & \textbf{0.78} \\
SeeAct & 20.9 & 11.3 & 0.61 \\
LiteWebAgent & 38.0 & 9.0 & 0.74 \\
OpenWebAgent & 28.4 & 10.5 & 0.67 \\
\bottomrule
\end{tabular}
}

\end{table}

Table~\ref{tab:web-agent-comparison} covers ten representative systems: general-purpose multimodal models under shared prompting, and specialized agents with vision-based, DOM-based, or combined interfaces. This provides complementary comparisons rather than an exhaustive survey.

\subsubsection{General-purpose LMMs.}
GPT-o1, GPT-4-Turbo, GPT-4o, GPT-4o-mini, and the Gemini~\cite{gemini2024} 1.5-pro, 2.0, and 2.5 models receive natural-language instructions and webpage screenshots. All use the same prompting format and temperature 0.1.

\subsubsection{Specialized Web Agents.}
\textbf{SeeAct}~\cite{zheng2024seeact} is a vision-based agent with perception and action-decoding modules; \textbf{LiteWebAgent}~\cite{litewebagent2025} parses DOM structures and predicts high-level actions; and \textbf{OpenWebAgent}~\cite{iong2024openwebagent} combines DOM modeling, visual grounding, tool use, retrieval-augmented inputs, and explicit memory.

\subsection{Evaluation Protocol}
Each of the ten systems is evaluated on all 510 tasks in three repeated runs to account for generation stochasticity. Success requires strict matching against the verified gold answer. Overall and category-level scores are the arithmetic means of the corresponding three run-level success rates.

Let $s_{ir}=1$ if task $i$ succeeds in run $r$, and $s_{ir}=0$ otherwise. The reported success rate is
\begin{equation}
\mathrm{SR} = \frac{100}{3N}\sum_{r=1}^{3}\sum_{i=1}^{N}s_{ir},
\label{eq:success-rate}
\end{equation}
where $N=510$ for the overall rate and $N$ is the category size for a category-level rate.

\subsection{Main Results}
Table~\ref{tab:WebArxiv-selected-models} shows clear model specialization. GPT-o1 achieves the best overall success rate (47.6\%), leading Website Info \& Org (73.3\%) and Paper Retrieval (66.7\%). Gemini-2.5 performs best on Rules \& Accounts (58.8\%) and Advanced Search \& Filters (46.3\%), while GPT-o1 leads Deep Paper Extraction (45.3\%), slightly above LiteWebAgent (44.3\%). Cross-Paper Comparison is lowest overall: the strongest base results are Gemini-2.5 (22.3\%), LiteWebAgent (21.3\%), and GPT-o1 (20.7\%).

Table~\ref{tab:WebArxiv-enhanced-comparison} compares base models with their dynamic-memory variants. Dynamic memory improves overall performance for five of six evaluated models: GPT-o1 with dynamic memory reaches 53.6\%, and Gemini-2.5 reaches 49.9\%, up from 44.4\%. Category-level effects include both gains and declines.

\begin{table}[!t]
\caption{Comparison of models with and without dynamic memory across six task categories.}
\label{tab:WebArxiv-enhanced-comparison}
\centering
\setlength{\tabcolsep}{2.8pt}
\renewcommand{\arraystretch}{0.98}
\resizebox{\linewidth}{!}{
\begin{tabular}{lcccccc|c}
\toprule
Web Agents &
\begin{tabular}[c]{@{}c@{}}Website\\ Info \& Org\end{tabular} &
\begin{tabular}[c]{@{}c@{}}Rules \&\\ Accounts\end{tabular} &
\begin{tabular}[c]{@{}c@{}}Paper\\ Retrieval\end{tabular} &
\begin{tabular}[c]{@{}c@{}}Adv. Search\\ \& Filters\end{tabular} &
\begin{tabular}[c]{@{}c@{}}Deep Paper\\ Extraction\end{tabular} &
\begin{tabular}[c]{@{}c@{}}Cross-Paper\\ Comparison\end{tabular} &
Total (\%) \\
\midrule
GPT-4-Turbo & 44.2\% & 33.9\% & 47.3\% & 25.3\% & 31.3\% & 14.7\% & 31.7\% \\
GPT-4-Turbo + dynamic memory & 52.7\% & 42.4\% & 46.3\% & 30.0\% & 29.0\% & 14.7\% & 33.8\% \\
\midrule
GPT-4o & 35.8\% & 29.7\% & 36.3\% & 24.3\% & 38.7\% & 17.7\% & 30.0\% \\
GPT-4o + dynamic memory & 63.6\% & 60.0\% & 38.3\% & 34.7\% & 52.7\% & 26.0\% & 43.1\% \\
\midrule
GPT-o1 & 73.3\% & 50.3\% & 66.7\% & 42.3\% & 45.3\% & 20.7\% & 47.6\% \\
GPT-o1 + dynamic memory & 73.3\% & 55.8\% & 64.7\% & 52.7\% & 60.3\% & 24.7\% & 53.6\% \\
\midrule
GPT-4o-mini & 52.1\% & 49.7\% & 56.3\% & 28.3\% & 34.3\% & 14.0\% & 37.1\% \\
GPT-4o-mini + dynamic memory & 57.6\% & 31.5\% & 52.7\% & 31.0\% & 35.7\% & 17.0\% & 36.3\% \\
\midrule
Gemini-1.5-pro & 47.9\% & 41.2\% & 53.3\% & 33.3\% & 38.3\% & 18.7\% & 37.8\% \\
Gemini-1.5-pro + dynamic memory & 60.0\% & 59.4\% & 51.7\% & 38.3\% & 45.7\% & 24.3\% & 44.2\% \\
\midrule
Gemini-2.5 & 63.0\% & 58.8\% & 54.3\% & 46.3\% & 36.3\% & 22.3\% & 44.4\% \\
Gemini-2.5 + dynamic memory & 81.8\% & 72.7\% & 56.3\% & 43.7\% & 41.0\% & 28.7\% & 49.9\% \\
\bottomrule
\end{tabular}
}

\end{table}

\begin{table}[!t]
\caption{Success rates under fixed-history and dynamic-memory strategies.}
\label{tab:arXiv-category-performance}
\centering
\setlength{\tabcolsep}{4pt}
\renewcommand{\arraystretch}{0.98}
\resizebox{0.72\linewidth}{!}{%
\begin{tabular}{lcc}
\toprule
Memory Mechanism & Successful ($\uparrow$) & Failed ($\downarrow$) \\
\midrule
GPT-4-Turbo last 3 steps & \textbf{31.7\%} & 68.3\% \\
GPT-4-Turbo last 2 steps & 30.5\% & 69.5\% \\
GPT-4-Turbo last step & 33.5\% & 66.5\% \\
GPT-4-Turbo + dynamic memory & \textbf{33.8\%} & 66.2\% \\
\midrule
GPT-o1 last 3 steps & \textbf{47.6\%} & 52.4\% \\
GPT-o1 last 2 steps & 50.8\% & 49.2\% \\
GPT-o1 last step & 52.5\% & 47.5\% \\
GPT-o1 + dynamic memory & \textbf{53.6\%} & 46.4\% \\
\bottomrule
\end{tabular}
}

\end{table}

\subsection{Ablation Studies}
We compare dynamic memory with fixed-window configurations for GPT-4-Turbo and GPT-o1. As shown in Table~\ref{tab:arXiv-category-performance}, GPT-o1 with dynamic memory reaches 53.6\%, outperforming the most recent step (52.5\%) and uniform three-step memory (47.6\%). Dynamic memory also improves GPT-4-Turbo from 31.7\% to 33.8\%, demonstrating the benefit of adaptive context retrieval and action generation.

\subsection{Analysis}
\begin{figure}[!t]
\centering
\begin{minipage}[t]{0.56\linewidth}
    \centering
    \includegraphics[width=\linewidth]{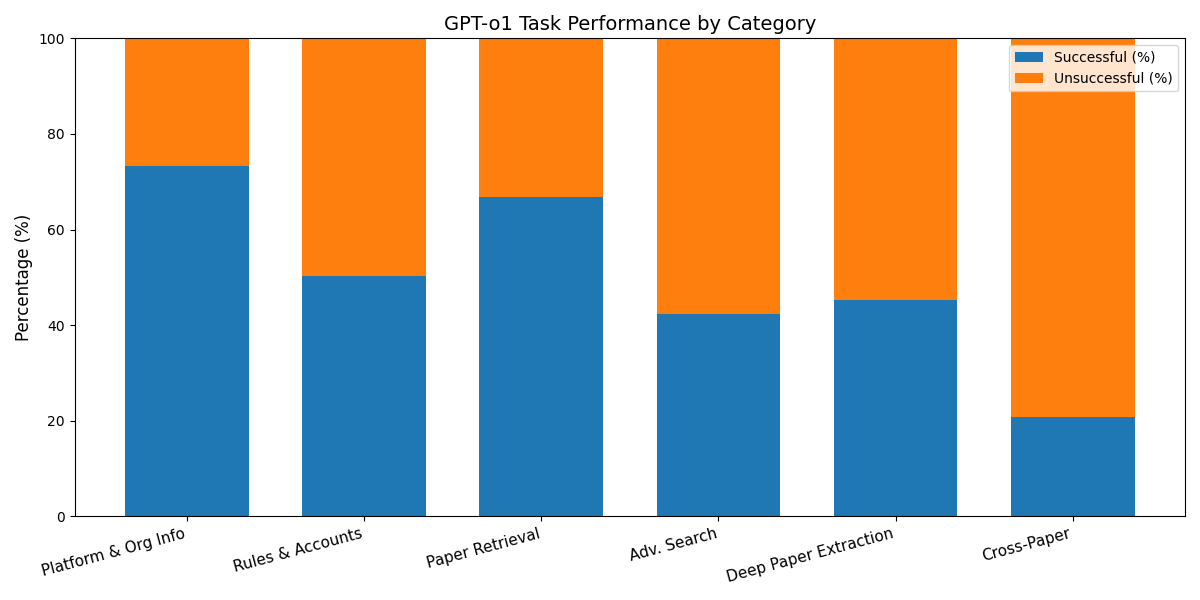}
    \caption{GPT-o1 completion rates by task category.}
    \label{fig:barchart}
\end{minipage}\hfill
\begin{minipage}[t]{0.40\linewidth}
    \centering
    \includegraphics[width=\linewidth]{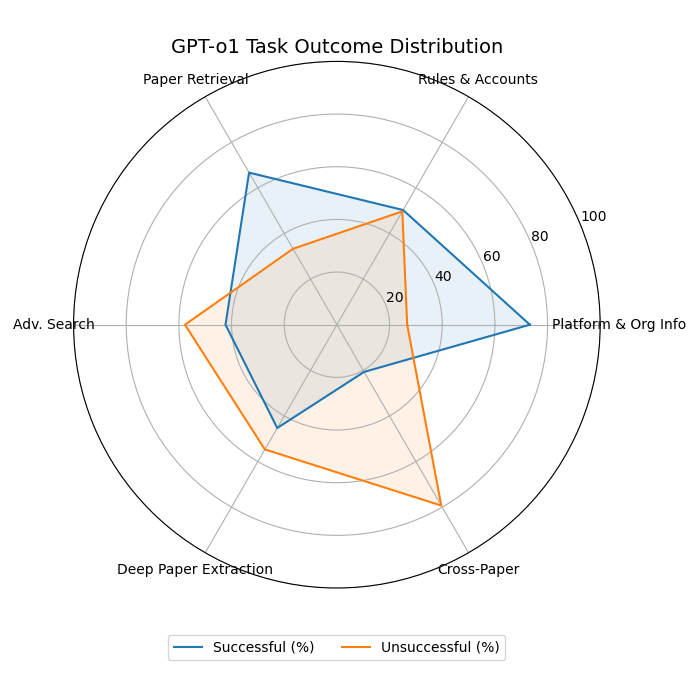}
    \caption{GPT-o1 success and failure rates by task category.}
    \label{fig:radarchart}
\end{minipage}
\end{figure}

\subsubsection{Cross-Category Performance.}
Figures~\ref{fig:barchart} and~\ref{fig:radarchart} show the difficulty gradient: Website Info \& Org and Paper Retrieval mainly involve lookup; Rules \& Accounts require exact policy statements, while Deep Paper Extraction requires navigation through long documents and HTML anchors. Advanced Search \& Filters combines Boolean operators, dates, and field constraints. Cross-Paper Comparison adds retrieval, alignment, and joint reasoning across documents and is the hardest category.

\subsubsection{Cross-Model Comparison.}
SeeAct underperforms across categories on these text-heavy interfaces. Model scale alone is not decisive: GPT-4o-mini remains competitive at 37.1\% overall. Model-specific category leaders are reported in the main results.

\subsubsection{Dynamic Memory Effects.}
Dynamic memory improves reported overall success for five of six models, but not every category (Table~\ref{tab:WebArxiv-enhanced-comparison}). Deep Paper Extraction shows substantial gains for GPT-o1 and GPT-4o, while GPT-4o-mini slightly declines overall. The category results therefore indicate model-dependent benefits from retrieving relevant earlier interactions.

\subsubsection{Error Analysis.}
Trajectory inspection reveals fixation on recent but irrelevant history and rigid attention patterns. Strict matching penalizes semantically correct paraphrases, especially for policy tasks. Field and anchor errors also occur on advanced-search or HTML-dense pages, where agents confuse title and abstract fields or miss a checkbox or link.

These findings carry three design implications: lightweight DOM priors for anchor-dense pages; selective rather than uniformly long memory; and constrained action templates for Boolean operators and structured-field assignments.

\section{Conclusion}
We introduced WebArxiv, a reproducible benchmark of time-invariant arXiv tasks that capture realistic scholarly interactions. Experiments evaluate general-purpose LMMs and specialized web agents and reveal over-reliance on fixed histories. The lightweight dynamic-memory mechanism adaptively retrieves relevant context during navigation and improves performance across multiple categories, with the largest relative gains reported for cross-paper comparison. WebArxiv provides a reproducible foundation for web-agent evaluation and for developing more adaptive agents.

\begin{credits}
\subsubsection{\ackname}
We thank Ms Wanqi Yang for her invaluable advice and support during the early stage of this research.

\subsubsection{Disclosure of Interests.}
The authors declare no competing interests.
\end{credits}

\bibliographystyle{splncs04}
\bibliography{custom}

\clearpage
\setcounter{webarxivAppendixStart}{\value{page}}
\appendix
\renewcommand{\theHsection}{appendix.\Alph{section}}
\section{Appendix}
\subsection{Task 51: Instruction and Interaction Trajectory}
\label{app:task}

\noindent\textbf{Task ArXiv--51.} ``Provide the name of the university publishing in this paper: There is no polynomial formula for the catenary and the tame degree of finitely generated monoids.''

\noindent\textbf{Reference answer.} University of Graz.

\begin{figure}[H]
\centering
\begin{tabular}{@{}p{0.49\linewidth}@{\hspace{0.02\linewidth}}p{0.49\linewidth}@{}}
\TaskFiftyOneScreenshot{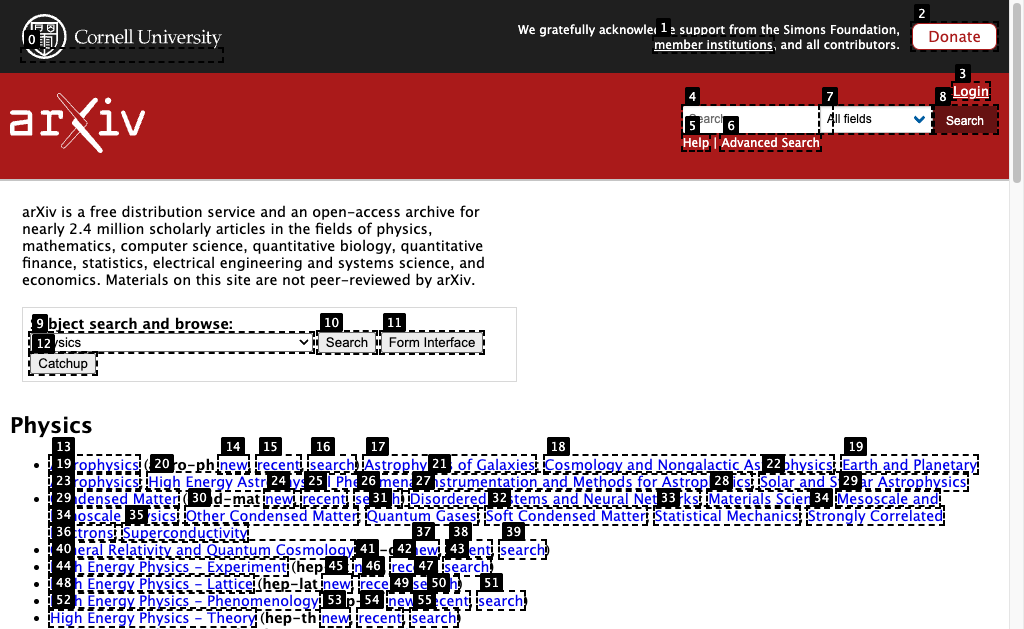} &
\TaskFiftyOneScreenshot{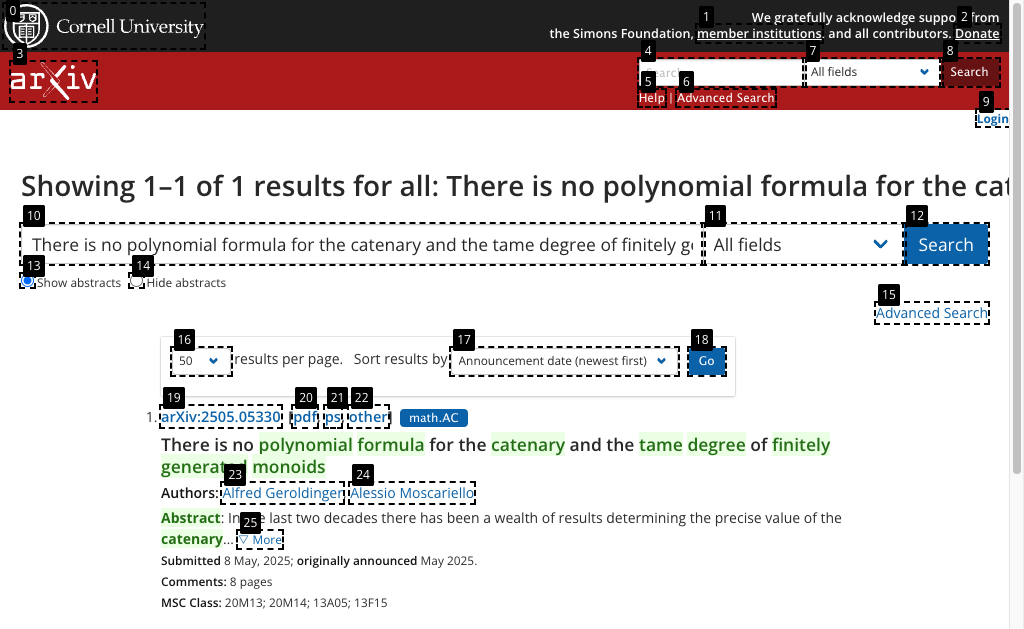} \\
\centering Step 1: Search paper & \centering\arraybackslash Step 2: Click [25] \\
\TaskFiftyOneScreenshot{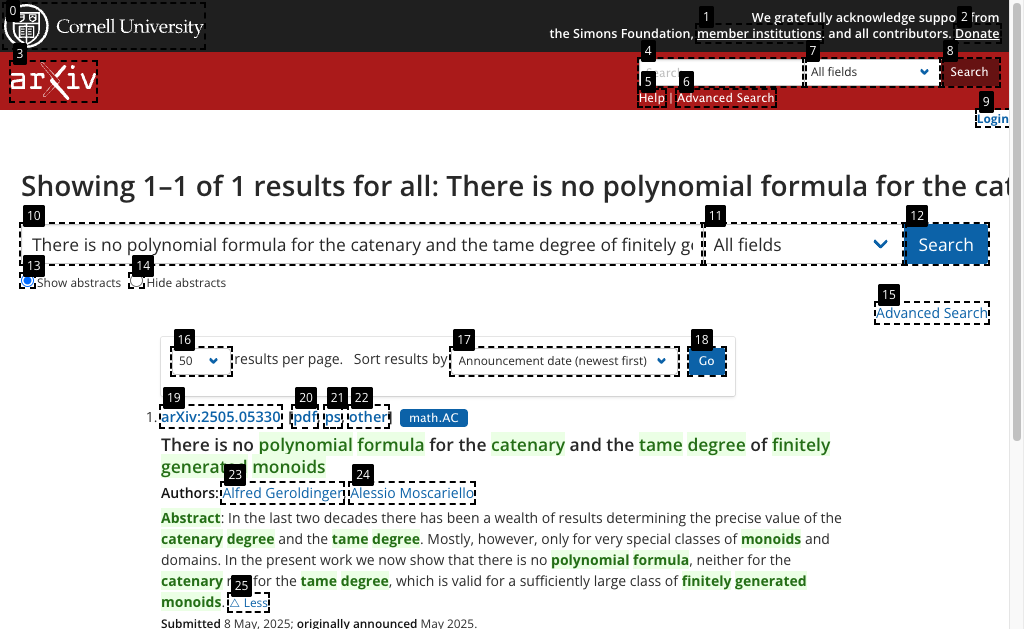} &
\TaskFiftyOneScreenshot{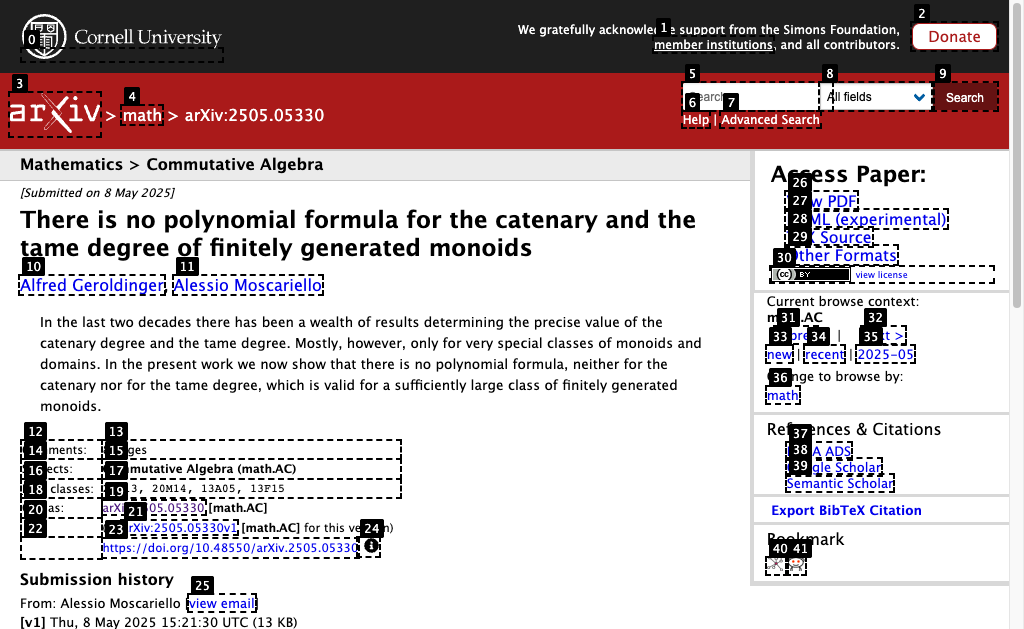} \\
\centering Step 3: Click [19] & \centering\arraybackslash Step 4: Click [27] \\
\TaskFiftyOneScreenshot{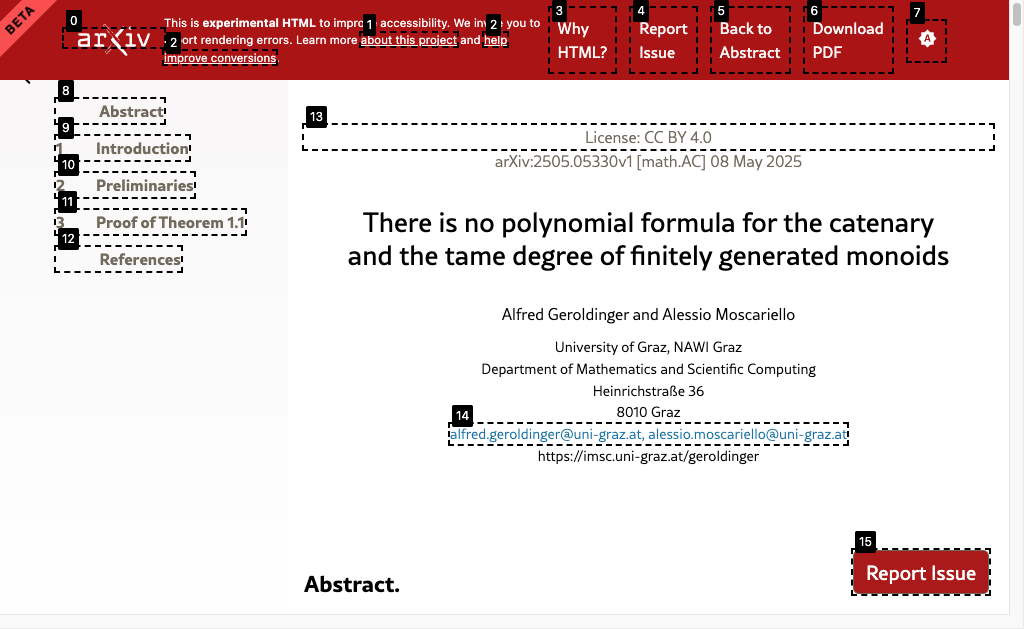} & \\
\centering Step 5: ANSWER & \\
\end{tabular}
\caption{The five screenshots for Task ArXiv--51. The agent searches for the specified paper, expands its metadata, opens the article detail page and HTML version, and returns ``University of Graz.'' The actions and reasoning are recorded in Table~\ref{tab:arxiv51-trajectory}.}
\label{fig:task51}
\end{figure}

Table~\ref{tab:arxiv51-trajectory} records the five-step interaction trace. Steps~1--5 correspond to the screenshots in Fig.~\ref{fig:task51}.

\begin{table}[H]
\caption{Interaction trajectory for Task ArXiv--51, illustrating targeted recall across a dense interface.}
\label{tab:arxiv51-trajectory}
\centering
\small
\setlength{\tabcolsep}{3pt}
\begin{tabularx}{\linewidth}{@{}>{\raggedright\arraybackslash}p{0.06\linewidth}>{\raggedright\arraybackslash}p{0.38\linewidth}>{\raggedright\arraybackslash}p{0.16\linewidth}>{\raggedright\arraybackslash}X@{}}
\toprule
Step & Thought (with dynamic memory) & Action & Action target (button/text) \\
\midrule
1 & To find the university, I should search the paper by its title. (Initial step---no prior context to reflect on.) & Type [4] & Search bar [4]; input: ``There is no polynomial formula for the catenary and the tame degree of finitely generated monoids'' \\
\midrule
2 & Search results are shown. From Step 1, I confirmed this is the correct paper. Let's open more details. & Click [25] & More [25]; expands metadata and abstract \\
\midrule
3 & (Dynamic memory: Step 2 showed author names, but no affiliations.) To get structured metadata, we should access the article detail page. & Click [19] & arXiv:2505.05330 [19]; opens paper's detail page \\
\midrule
4 & (Dynamic memory: Step 3 showed abstract, still no affiliations.) The HTML version might include full author info and contact details. & Click [27] & HTML (experimental) [27]; opens the full paper in HTML format \\
\midrule
5 & (Dynamic memory: Step 4 gave full text. University found in both email domain and author line.) & ANSWER; University of Graz & From author info: the \texttt{uni-graz.at} domain; full text shows: University of Graz \\
\bottomrule
\end{tabularx}
\end{table}

\clearpage
\subsection{Task 51: Screenshot Trajectory}
\label{app:task51-screenshots}

\subsubsection{Generative AI Use Disclosure.}
Generative AI tools, including OpenAI Codex, were used for language editing, \LaTeX{} formatting, and assistance auditing and repairing code, documentation, and the structural formatting of supplementary data files. They were not used to generate or validate scientific results or reference answers. The authors reviewed the resulting changes and take responsibility for all scientific content, data, results, references, and released code.

\clearpage
\setcounter{webarxivAppendixPages}{\numexpr\value{page}-\value{webarxivAppendixStart}\relax}
\setcounter{webarxivTotalPages}{\numexpr\value{page}-1\relax}
\setcounter{webarxivMainPages}{\numexpr\value{webarxivTotalPages}-\value{webarxivAppendixPages}\relax}
\typeout{WEBARXIV PAGE CHECK: main including references = \arabic{webarxivMainPages}; appendix = \arabic{webarxivAppendixPages}; total = \arabic{webarxivTotalPages}.}
\ifnum\value{webarxivMainPages}=12\relax
  \typeout{WEBARXIV MAIN PAGE CHECK: PASS (exactly 12 pages).}
\else
  \PackageWarningNoLine{WebArxiv}{Main content including references occupies \arabic{webarxivMainPages} pages; AJCAI requires exactly 12. Check the final compiled PDF}
\fi
\ifnum\value{webarxivAppendixPages}>2\relax
  \PackageWarningNoLine{WebArxiv}{Appendix occupies \arabic{webarxivAppendixPages} pages; the maximum is 2. Check the final compiled PDF}
\else
  \typeout{WEBARXIV APPENDIX PAGE CHECK: PASS (at most 2 pages).}
\fi

\end{document}